\documentclass{revtex4}
\usepackage{color}
\usepackage{amsmath}
\usepackage{graphicx}
\usepackage{array}
\usepackage{siunitx}

\begin{document}

\title{Group Delay Control in Longitudinal Offset Coupled Resonator Optical
Waveguides}
\author{Pedro~Chamorro-Posada}
\affiliation{Departmento de Teor\'{\i}a de la Se\~nal y Comunicaciones e Ingenier\'{\i}a
Telem\'atica, Universidad de Valladolid, ETSI Telecomunicaci\'on, Paseo
Bel\'en 15, E-47011 Valladolid, Spain.}
\author{F. Javier Fraile-Pel\'aez}
\affiliation{h Dept. Teor\'{\i}a de la Se\~nal y Comunicaciones, Universidad de Vigo, EI
Telecomunicaci\'on, Campus Universitario, E-36310 Vigo (Pontevedra), Spain}

\begin{abstract}

We study the differential group delay that can be obtained in coupled micro-ring resonator waveguides with longitudinal offset couplers.  Various devices based on this effect are proposed and analyzed using a finite-differences time-domain method and their potential applications are discussed.

\end{abstract}

\maketitle

\section{Introduction}

  The recent advances in fabrication technologies have promoted the microring resonator devices to the status of preferred platform for many photonic signal processing functionalities, such as optical filtering, dispersion compensation, sensing, or group delay control \cite{Lenz,jlt,Madsen,Chao,Boyd,Pedro1,Heebner,Poon}. Although microring structures can be realized using different materials, most proposals of passive structures have focused on silicon-on-insulator (SOI) devices to date, due to obvious reasons of convenience and the available enabling technology.

    In designing photonic structures at the top-level, a waveguide coupler is most conveniently treated as an ideal, lumped component characterized by the numerical value of its coupling constant.  However, in the standard fabrication process, the coupling constant is determined mainly by the transversal offset, i.e. by the separation between the coupler waveguides. To meet the accuracy typically demanded for the value of the coupling constant, such separation must be set with a precision of around a few nanometers, which is near the current technological limit.

    To overcome the difficulty described above, an alternative approach was proposed in \cite{domenech2009,domenech2011}. This technique consists in varying the coupling length rather than the inter-guide separation, which remains constant. Such variation is in turn obtained by changing the longitudinal offset between the two parallel coupled straight waveguides of fixed length. Photolitographic production is then feasible with this technique as the resolution requirements are strongly relaxed.

    As it stands, the longitudinal offset technique unavoidably imposes a certain type of phase imbalance between the fields at the two outputs of any coupling section. Thus, relative phase-shifts appear and accumulate throughout the whole structure.  In this work, we study the control of the relative group delay that can be attained and its use in various photonic devices.  A proof of principle of the operation of these designs is obtained by the simulation of their performance using a finite-difference time-domain (FDTD) solver for 2D Maxwell equations.  {{Applications of the proposed devices in optical beamforming networks and quantum information systems are discussed.}}

\begin{figure}[tbp]
\centering
\includegraphics[width=10cm]{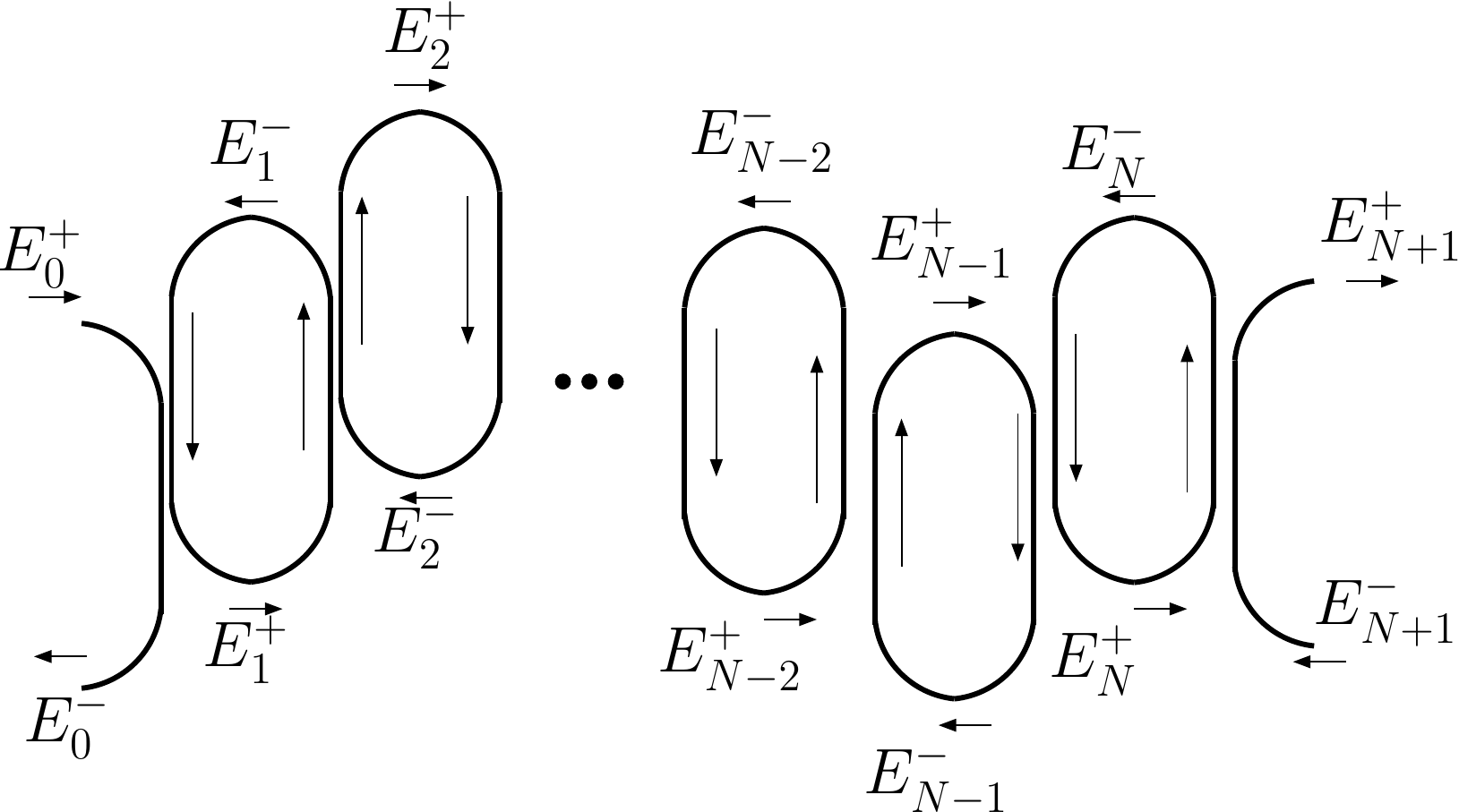}
\caption{Geometry of a coupled resonator optical waveguide implementing the
longitudinal offset technique.}
\label{fig::crow}
\end{figure}

\section{Model equations}

\begin{figure}[tbp]
\centering
\includegraphics[width=5cm]{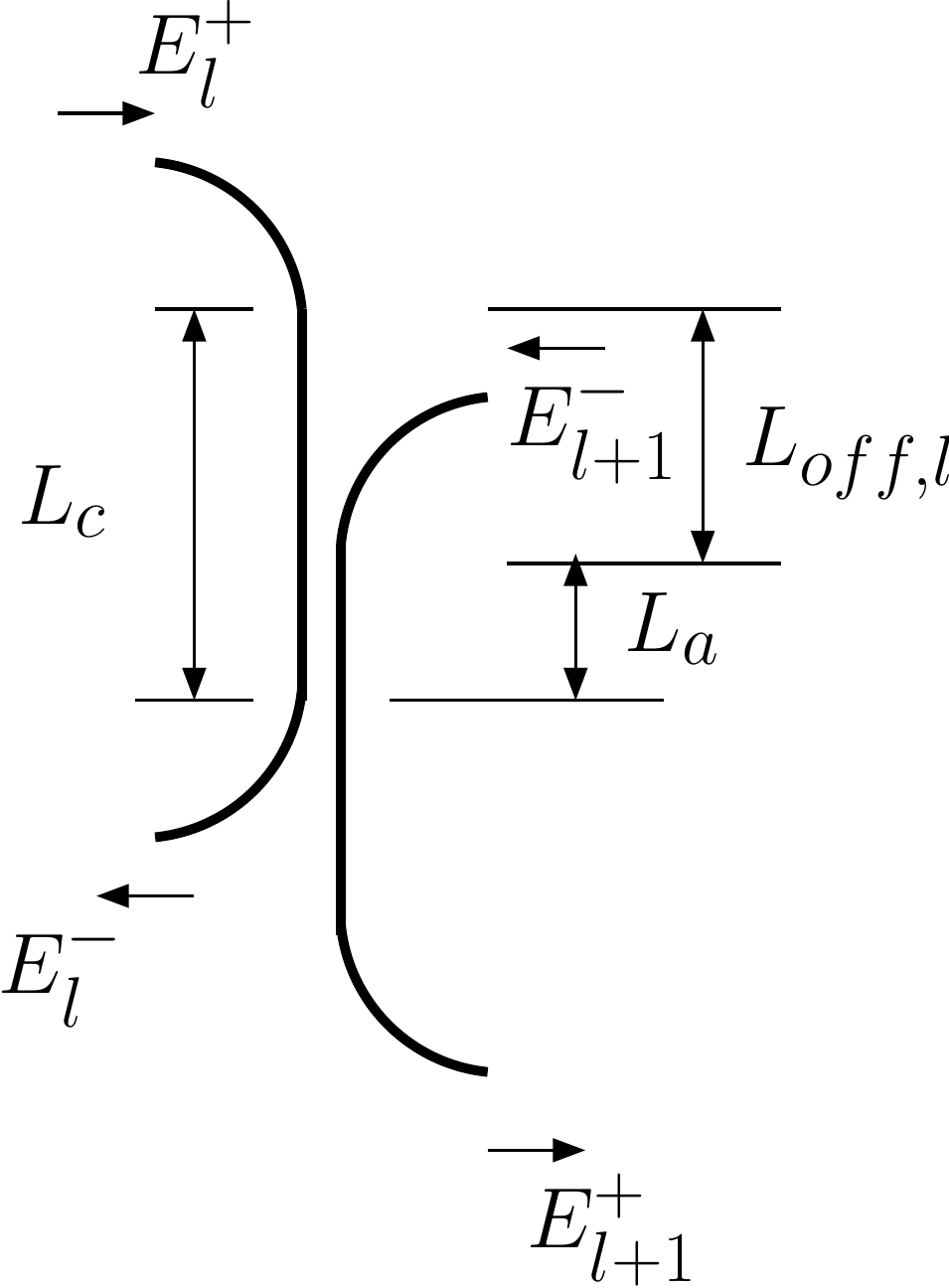}
\caption{Unit cell in a longitudinal offset coupled resonator optical
waveguide.}
\label{fig::celda}
\end{figure}

The geometry of a CROW implementing the longitudinal offset technique
is shown in Figure \ref{fig::crow}.  
With reference to Figure \ref{fig::celda}, $L_{c}$ is the length of the
straight waveguide sections and corresponds to the coupler length for zero
lateral displacement, $L_{R}=\pi R$, is the length of each curved waveguide
section of radius $R$. The half ring length is $L=L_{R}+L_{c}$ and will be
assumed constant. The longitudinal offset $L_{\text{off}}$ is defined in the
figure and will be regarded as positive whenever the displacement is along
the direction of the propagation of the field in that section and negative
otherwise.

We can trace the effect of the coupling and the propagation phase in Figure %
\ref{fig::celda} to write the field relations 
\begin{eqnarray}
E_{l+1}^{+} =-jt_l E_{l}^{+}\exp (-j\beta (L+L_{\text{off},l}))+r_l E_{l+1}^{-}%
\exp (-j\beta L)  \notag \\
E_{l}^{-} =r_l E_{l}^{+}\exp (-j\beta L)-jt_l E_{l+1}^{-}\exp (-j\beta (L-L_{%
\text{off,}l})).
\end{eqnarray}

Rewriting the above expressions as 
\begin{eqnarray}
E_{l}^{+} =\dfrac{1}{jt_l}\left( -E_{l+1}^{+}\exp (j\beta (L+L_{\text{off,}%
l})+r_l E_{l+1}^{-}\exp (j\beta L_{\text{off,}l})\right)   \notag \\
E_{l}^{-} =\dfrac{1}{jt_l}\left( -r_l E_{l+1}^{+}\exp (j\beta L_{\text{off,}%
l})+E_{l+1}^{-}\exp (-j\beta (L-L_{\text{off,}l}))\right) 
\end{eqnarray}%
permits to obtain the one-cell matrix relation 
\begin{equation}
\left( 
\begin{matrix}
E_{l}^{+} \\ 
E_{l}^{-}%
\end{matrix}%
\right) =\dfrac{1}{jt_{l}}\exp \left({j\beta L_{\text{off},l}}\right)
\left( 
\begin{matrix}
-\exp (j\beta L) & r_{l} \\ 
-r_{l} & \exp (-j\beta L)%
\end{matrix}%
\right) \left( 
\begin{matrix}
E_{l+1}^{+} \\ 
E_{l+1}^{-}%
\end{matrix}%
\right) .
\end{equation}%
and the input-output maxtrix relation for the whole chain 
\begin{equation}
\left( 
\begin{matrix}
E_{0}^{+} \\ 
E_{0}^{-}%
\end{matrix}%
\right) =\exp \left( j\beta \sum\limits_{l=0}^NL_{\text{off,}%
l}\right)
\left[
\prod\limits_{l=0}^{N}
\dfrac{1}{(jt_l)}
\left( 
\begin{matrix}
-\exp (j\beta L) & r_{l} \\ 
-r_{l} & \exp (-j\beta L)%
\end{matrix}%
\right) 
\right]
\left( 
\begin{matrix}
E_{N+1}^{+} \\ 
E_{N+1}^{-}%
\end{matrix}%
\right) .  \label{evolucion}
\end{equation}

 If the input signal is placed at $E_0^+$, taking into account that $%
E_{N+1}^-=0$, we can obtain $E_{N+1}^+=1/m_{11}E_o^+$, where $m_{11}$ is the
top diagonal element in the matrix relation \eqref{evolucion}.

Expression \eqref{evolucion} evidences that, for the specific case of a CROW 
chain and within the scope of the ideal model (wavelength-independent coupling constants), 
the offset-induced phase imbalance is harmless as far as the amplitude response is concerned, 
which was misinterpreted in \cite{phas}, while a contribution to the
group delay that is the cummulative effect of the longitudinal 
offsets appears in the system response. This effect can be engineered to yield added functionalities to the structure, as described below.

\begin{figure}[tbp]
\centering
\includegraphics[width=8cm]{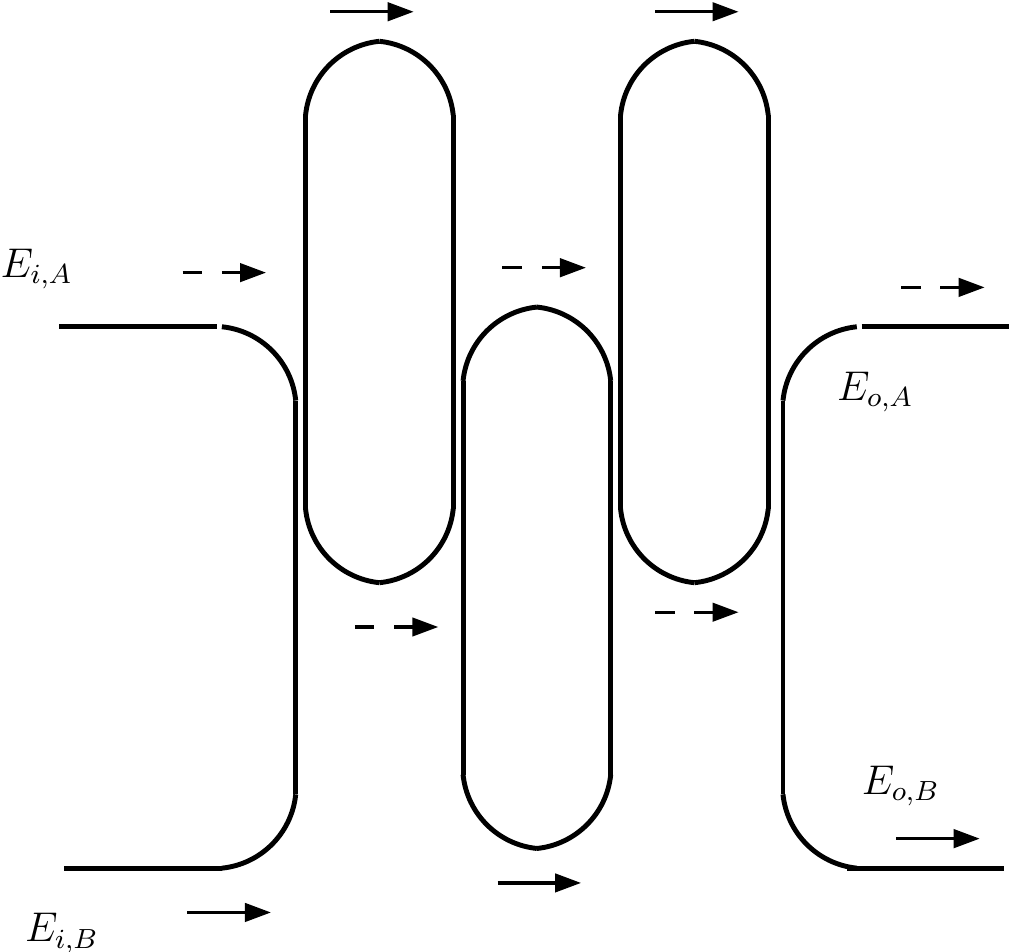}
\caption{The two  transfer functions implemented in a longitudinal offset CROW. }
\label{fig::esquema}
\end{figure}

The geometry depicted in Figure \ref{fig::esquema} shows that there are two sets of input/output ports available, which are  respectively marked as $A$ and $B$. 
When the transmission path is changed from $A$ to $B$, there is a sign flip for every value of $L_{\text{off}}$ along the signal path.  According to Equation \eqref{evolucion}, this implies a net change in the group
delay when the transmission topology is changed. Therefore, there exist two possible transfer functions implemented by the same physical system that are 
identical except for a global group delay term that can take two opposite values $\tau _{B}=-\tau _{A}=n_g/c\sum_{l=0}^N L_{\text{off,}l}\equiv \tau_{\text{off}}$.  

Figure \ref{fig::esquema} displays the schematic of a $N=3$ ring CROW.  The transmission path for the \emph{fast} $A$ input-output ports is marked with dashed arrows and the \emph{slow} path relating the $B$ input-output ports is marked with solid arrows.   The transfer functions are 
\begin{eqnarray}
H_A(\omega)=\exp(j \tau_{\text{off}} \omega) H(\omega) \nonumber\\
H_B(\omega)=\exp(-j \tau_{\text{off}} \omega) H(\omega),\label{eq::FT}
\end{eqnarray}  
where $H(\omega)$ is a common term that would correspond to the transfer function of the same system implemented without the longitudinal offsets.

\begin{figure}[tbp]
\centering
\includegraphics[width=10cm]{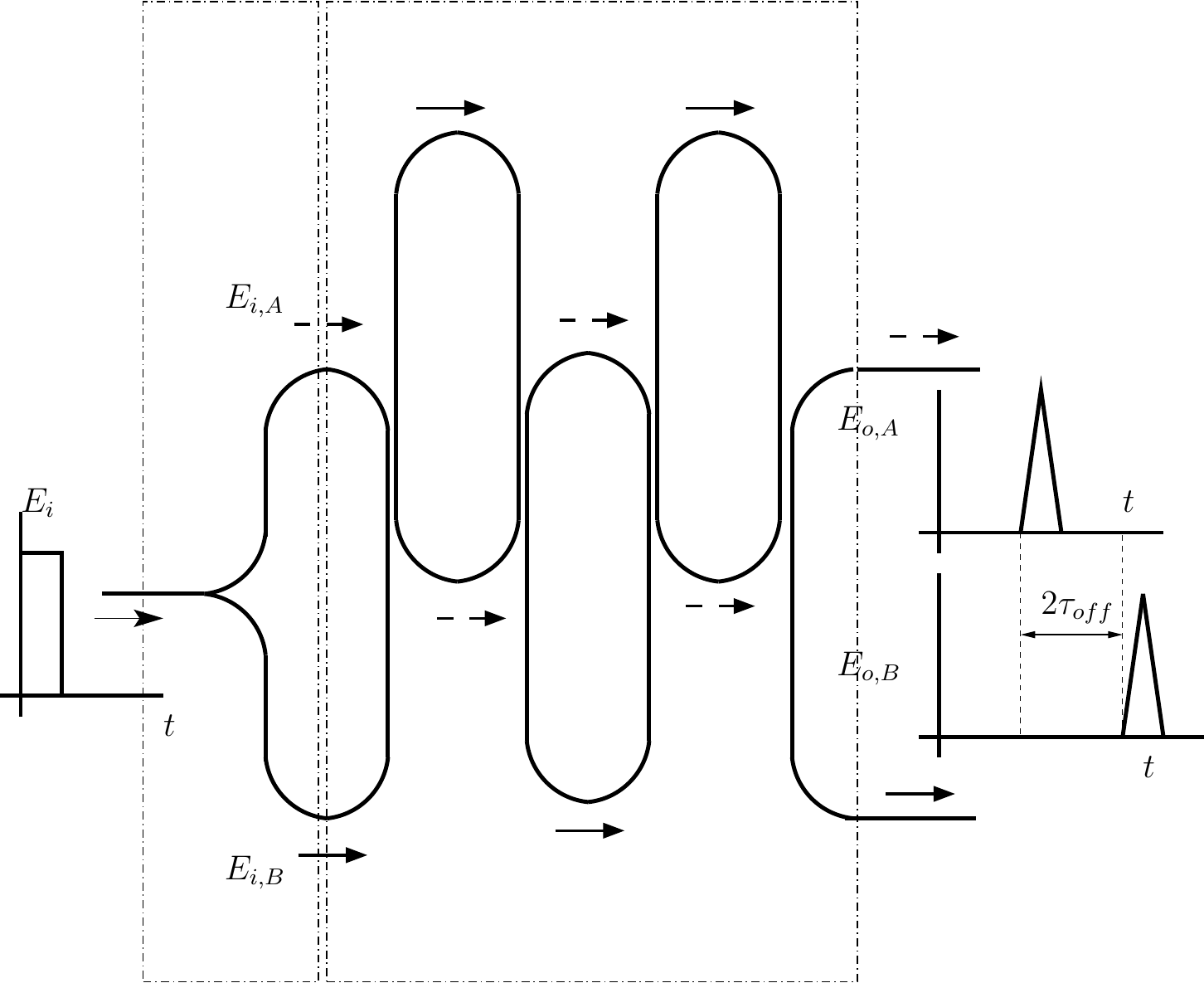}
\caption{Structure with two output replicae with a controlled relative group delay.}
\label{fig::dossalidas}
\end{figure}

\begin{figure}[tbp]
\centering
\includegraphics[width=10cm]{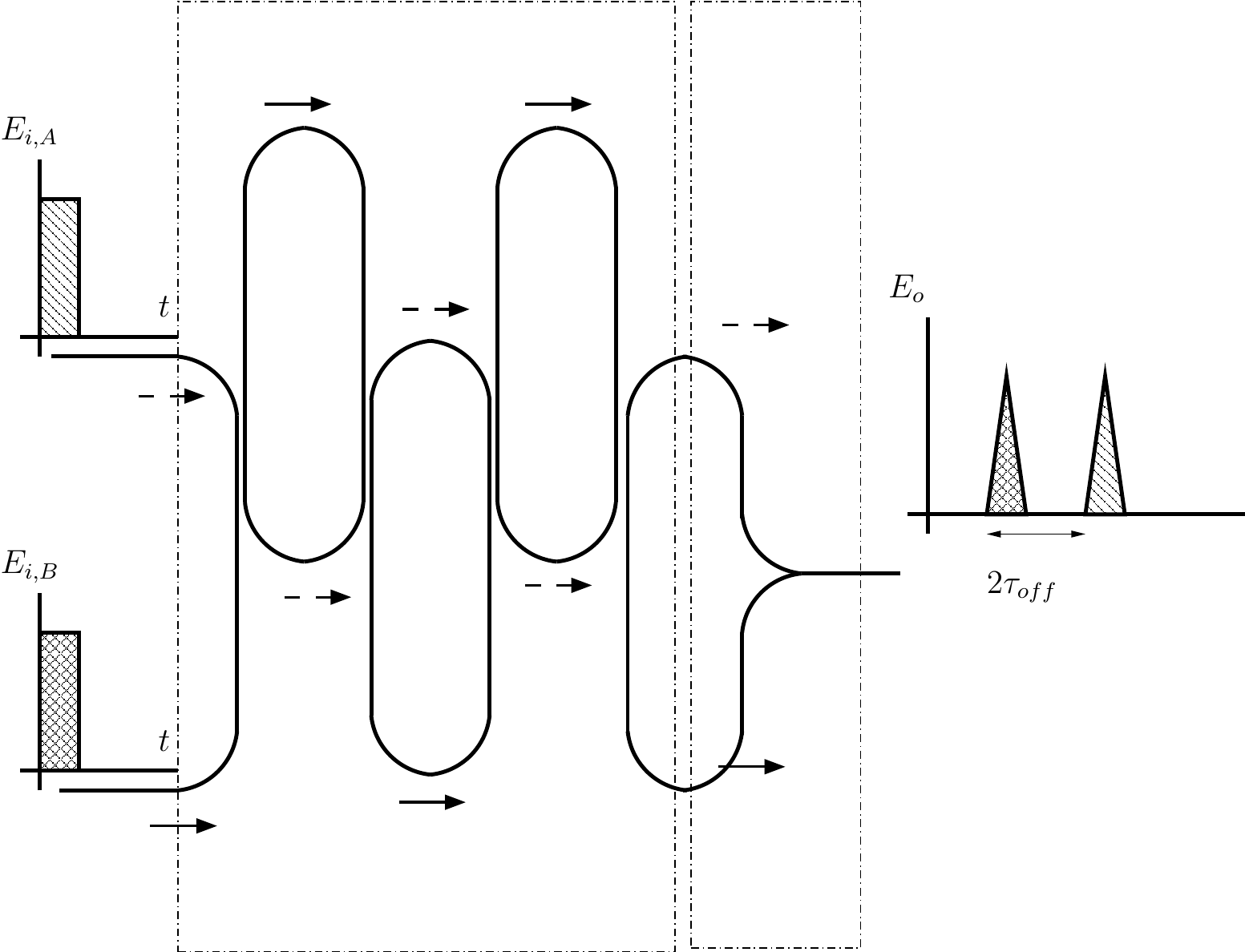}
\caption{Time-domain multiplexer.}
\label{fig::tdm}
\end{figure}

\begin{figure}[tbp]
\centering
\includegraphics[width=10cm]{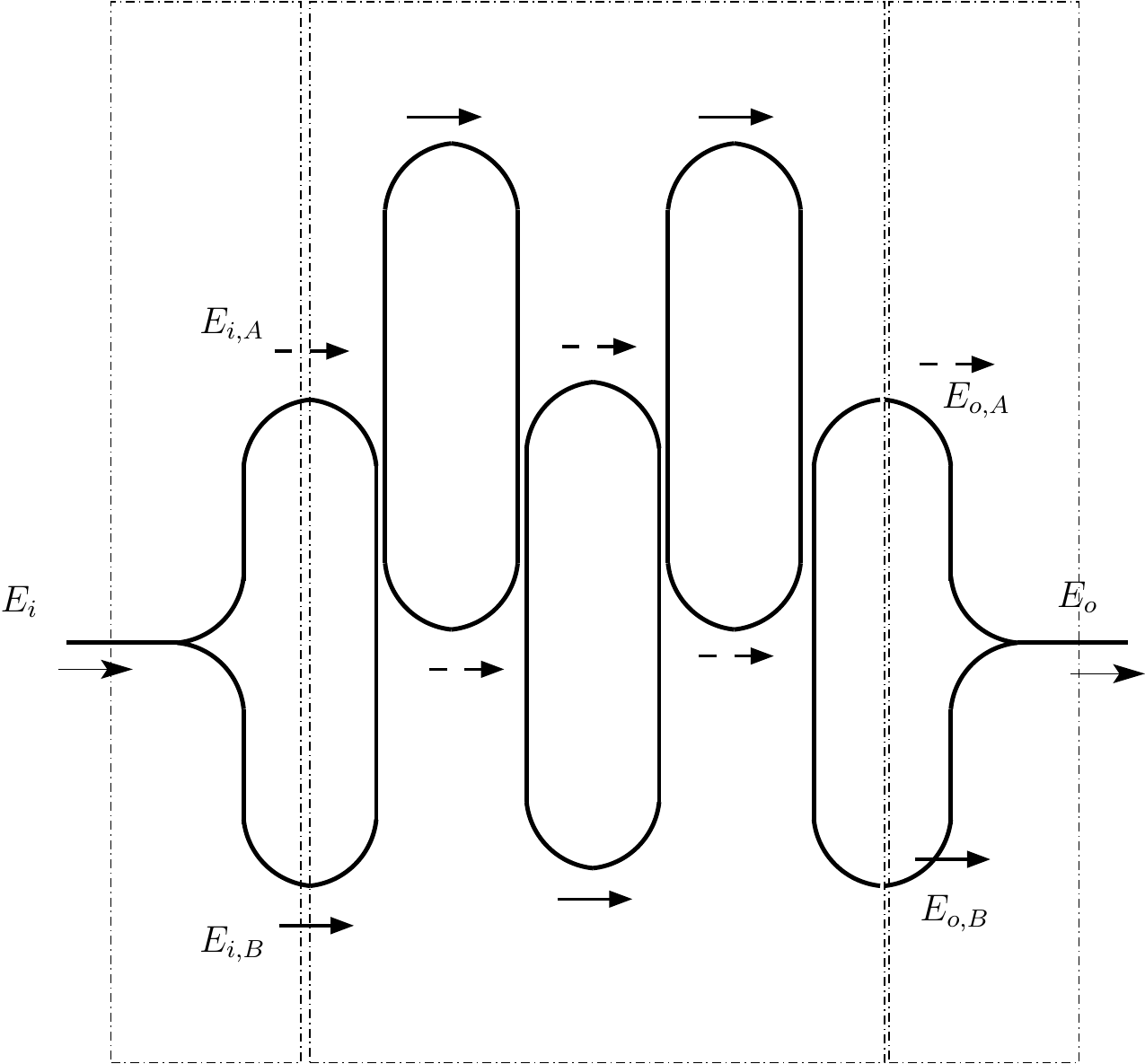}
\caption{Mach-Zehnder implementation using a longitudinal displacement
structure.}
\label{fig::mz}
\end{figure}

\begin{figure}[tbp]
\centering
\includegraphics[width=4cm]{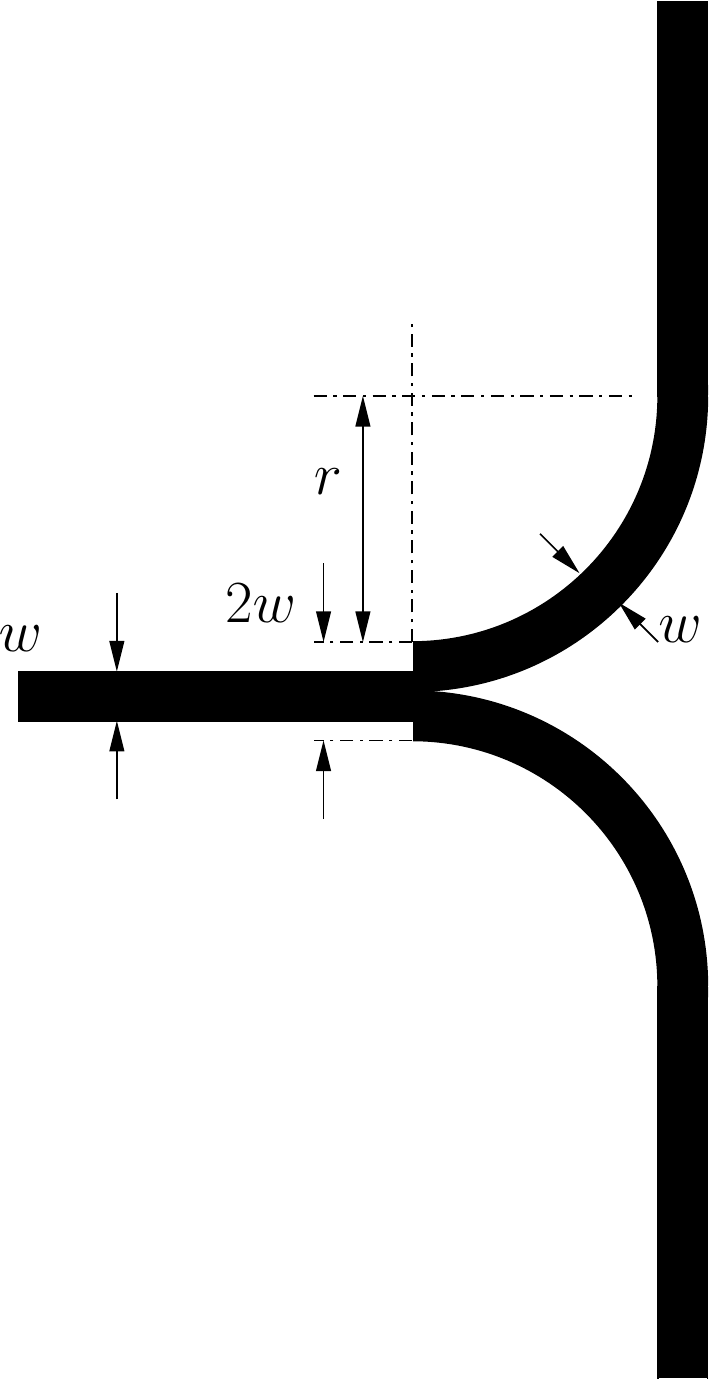} 
\caption{Geometry of the coupler/splitter implemented in the simulations.}
\label{fig::acoplador}
\end{figure}

\begin{figure}[tbp]
\centering
\begin{tabular}{c}
(a) \\ 
\includegraphics[width=8cm]{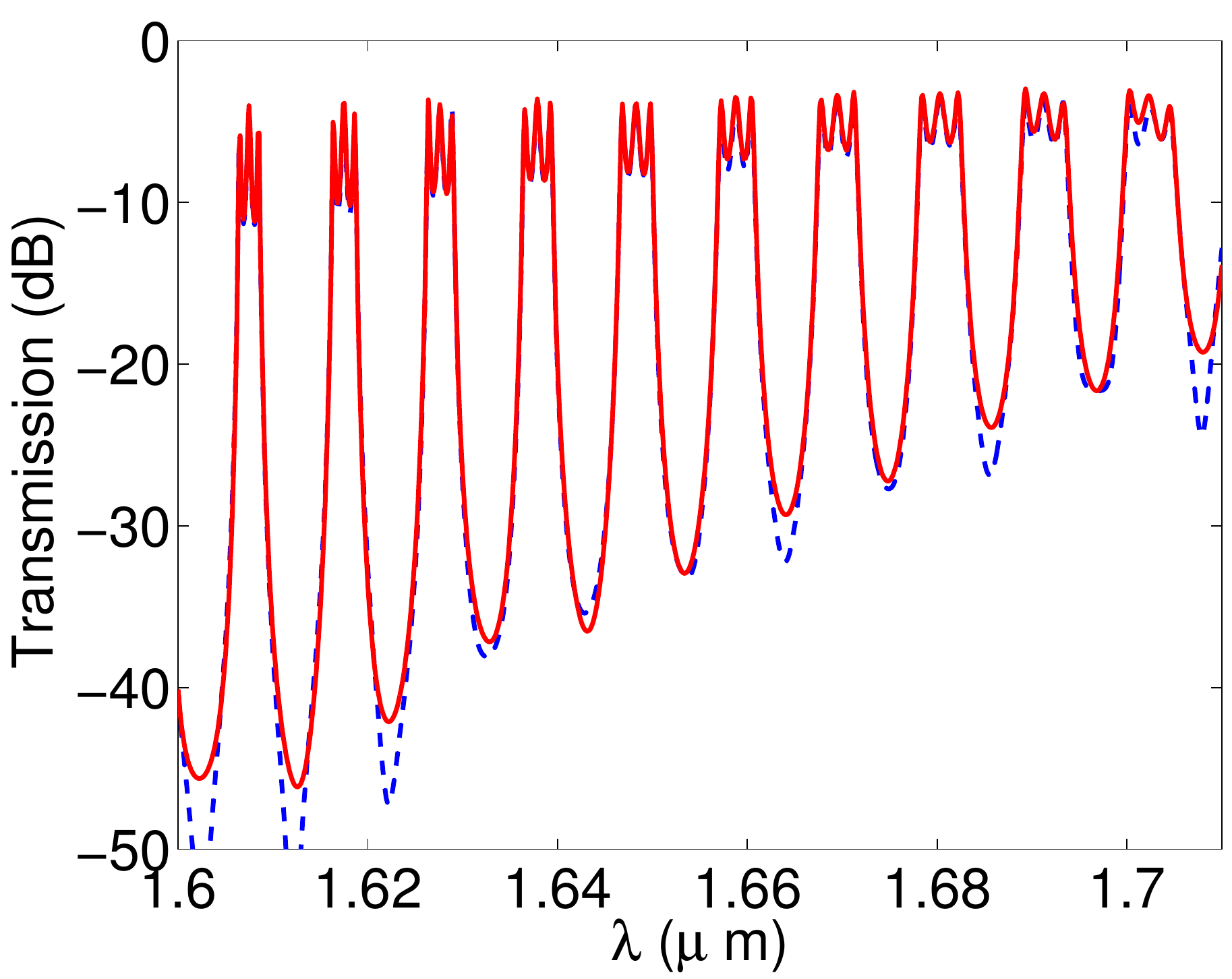}\\  %
(b)\\
\includegraphics[width=8cm]{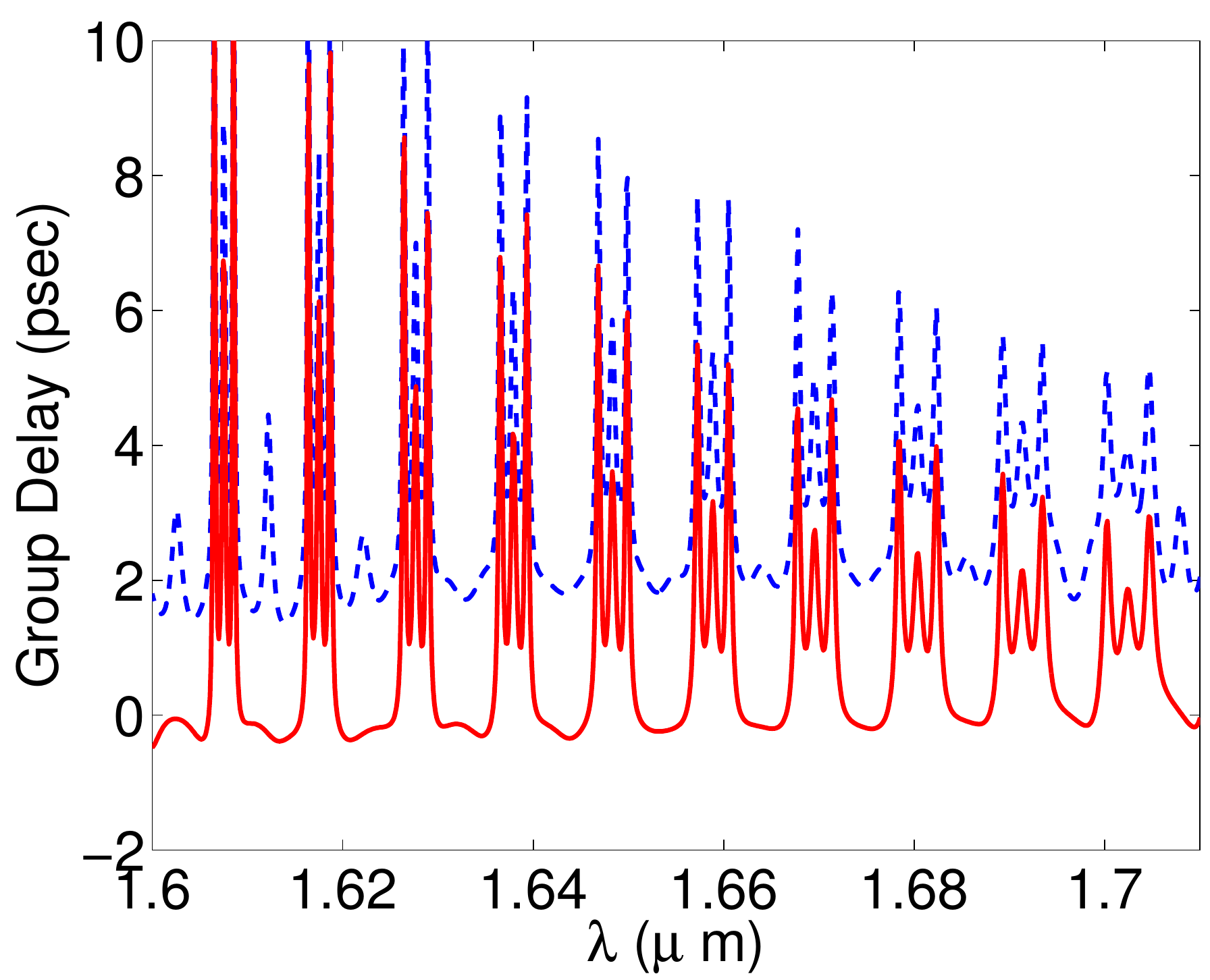}%
\end{tabular}%
\caption{Transmission spectrum (a) and group delays (b).  Fast input-output transfer function is indicated with a solid red line and the slow one with a blue dashed trace.}
\label{fig::espectros}
\end{figure}

\begin{figure}[tbp]
\centering
\includegraphics[width=8cm]{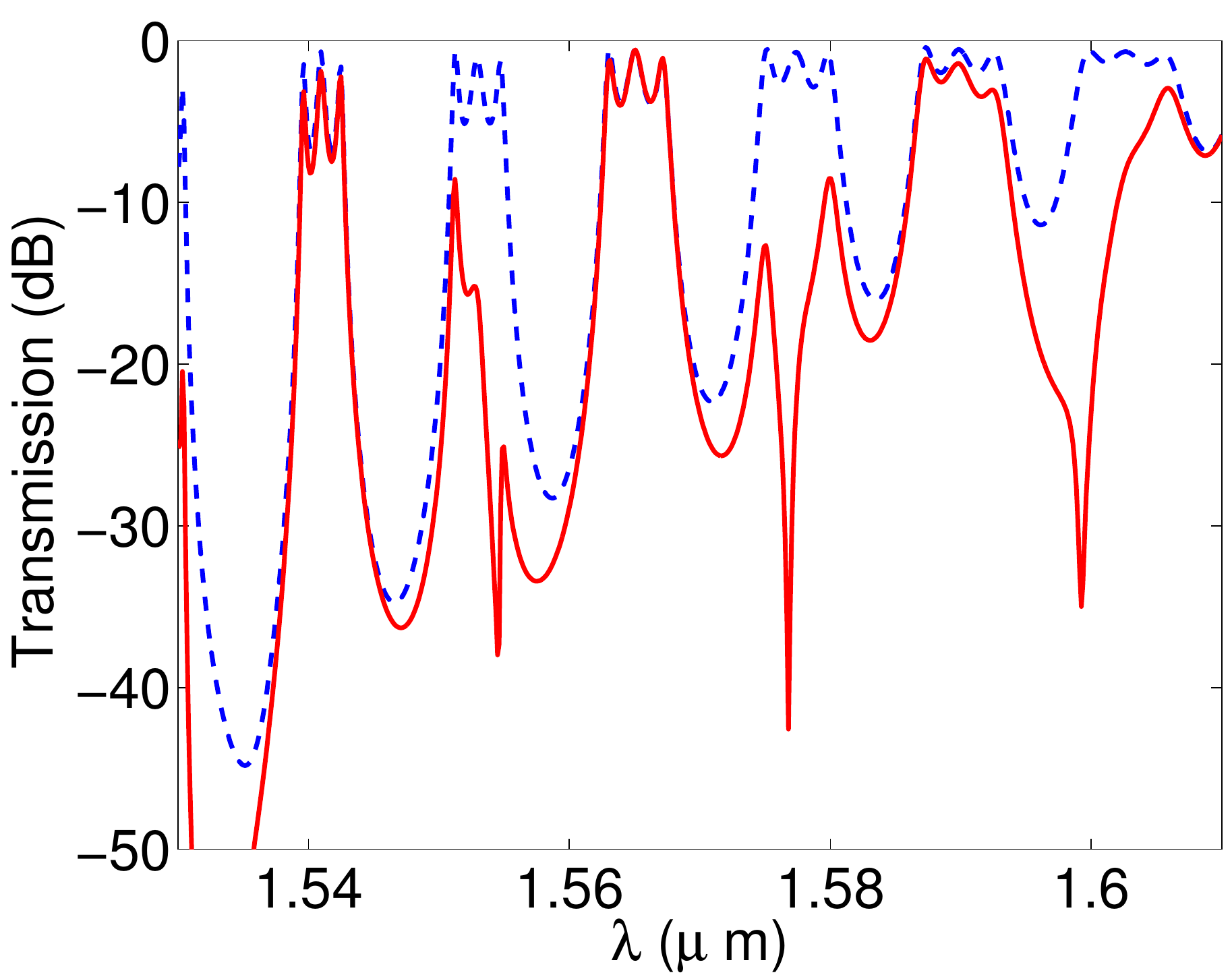} 
\caption{Transmission spectrum for a single input/output (dashed line) and with FSR doubling using the MZ configuration (solid line).}
\label{fig::espectromz}
\end{figure}

It should be noted that each input port ($A$ or $B$) coincides
with the reflection path for the converse configuration. When
both transmission paths are simultaneously used, this circumstance
will generally call for taking additional measures at the
design stage, depending on the specific application. However, even if the reflected signals cannot
in general be ignored, in this specific type of CROW filter the reflected spectral components
correspond mainly to the band-stop transmission of the filter
\cite{landobasa}, the reflection level thus being negligible when the transmitted signal is predominantly
localized in the filter pass-band, which avoids the problem. The reflected signals can also
be used to engineer complex architectures, as detailed below.

\section{Applications}

Using a Y-branch to combine both inputs, as shown in Figure \ref{fig::dossalidas}, we can
simultaneously obtain two outputs with the same amplitude spectral shaping and 
a relative delay. The shape used for the representation of the signals intends to 
reflect the existing relation in the filtering due to the common term $H(\omega)$ in
\eqref{eq::FT}.  {{Cascading this
type of structures  would permit to obtain a series of relatively delayed signals generated
from the same input with a given filtering.  The filtering in the first stage will substantially mitigate the effect of reflections in subsequent elements}}.
Reversing the scheme, as shown 
in Figure \ref{fig::tdm}, permits to multiplex two signals with a controlled 
relative time-delay and simultaneously apply the same filtering to both inputs. The relative delay between the two channels is given by the expression
\begin{equation}
\Delta\tau = 2\tau_{\text{off}}=2\dfrac{n_g}{c}\sum_{l=0}^N L_{\text{off,}l}.\label{eq::tb}
\end{equation}
For equal $L_{\text{off},l}=L_{\text{off}}$, the above expression reads $\Delta\tau=2(N+1)\dfrac{L_{\text{off}}n_g}{c}$.

The connection of the two inputs using a 3 dB splitter and the outputs with a 3 dB coupler, 
as shown in Fig. \ref{fig::mz}, leads to the
implementation of a Mach-Zehnder (MZ) interferometer and an overall transfer function
\begin{equation}
H_{MZ}(\omega)=\cos(\tau_{\text{\text{off}}}\omega)H(\omega).
\end{equation}
This type of response can be exploited to double the FSR of an optical filter by adequately tuning of
the relative delay.  The position of a spectral null every other transmission band of the CROW filter requires the condition $\tau=4\tau_{\text{off}}$ or, for constant $L_{\text{off},l}=L_{\text{off}}$, with
\begin{equation}
L_{\text{off}}=\dfrac{L}{2(N+1)}.
\end{equation}
Taking into account that $L=L_a+L_{\text{off}}+\pi R$, with $L_a$ defined in Fig. \ref{fig::celda}, we can write 
\begin{equation}
L_{\text{off}}=\dfrac{L_a+\pi R}{2N+1}.\label{eq::disenio1}
\end{equation}
A design based on the above condition \eqref{eq::disenio1} is too sensitive to small deviations of the microring parameters in the fabrication process.  A more robust design can be obtained by seeking the desired response only in a limited bandwidth. If  $\tau=s\tau_{\text{off}}$, with $s$ a rational number close to $4$, the MZ and CROW resonance conditions will coincide at a particular band of the CROW periodic response and an approximate cancellation will happen in a certain bandwidth that will depend on the value of $s$.  Fabrication tolerances in this case will affect the actual response of the filter, but a good approximation to the original design can be obtained in the presence of small deviations from the ideal design parameters.  In this case, the value of a constant offset for a given coupler length is given by the expression
\begin{equation}
L_{\text{off}}=\dfrac{2L_a+2\pi R}{(N+1)s-2}.\label{eq::disenio2}
\end{equation}

By adding a phase control to one the output branches, the MZ configuration can 
be used to produce and process quantum superpositions of time-bin states
for quantum information applications \cite{brendel,tittel}.  Instead of a design based on Equation \eqref{eq::disenio2}, as in the previous FSR doubling application,  the offsets are calculated such that \eqref{eq::tb} produces the required time-bin separation which has to be larger than the pulse width. This proposal permits an integrated optics implementation with a precise control of the
photon waveform through the filter spectral shaping and is better suited to relative delays in the picosecond range \cite{TB1,TB2}.

\begin{figure}[tbp]
\centering
\includegraphics[width=3.5in]{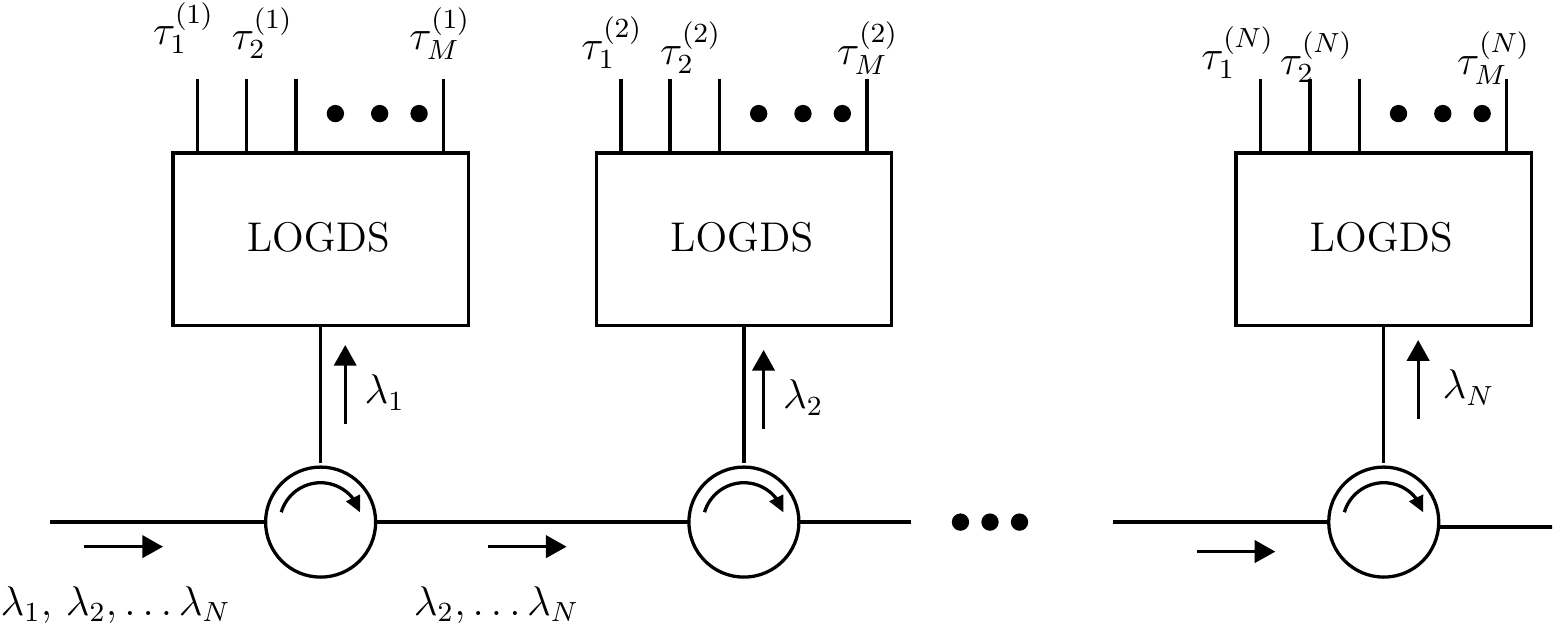} 
\caption{Architecture implementing multiple group delay structures in a WDM system.}
\label{fig::adddrop}
\end{figure}

{{We now consider a longitudinal offset group delay shifter (LOGDS) built by cascading various systems of the type described in Fig. \ref{fig::dossalidas}.  If the transmission bandwidth is small compared to the FSR,  multiple delayed outputs of the signal present in a single wavelength channel will be produced at the output, whereas most of the spectral content present in the optical input will be reflected.  This can be used for the realization of complex WDM architectures for generating multiple delayed signals at different channels, as illustrated in Figure \ref{fig::adddrop}, which could find application, for instance, in advanced optical beamforming networks \cite{zhuang}.  The frequency selectivity is strictly required only in the first stage of the LOGDS tree that can be implemented with a high $r$ single ring stage.  After the extraction of the drop channel, the rest of the WDM multiplex is reinjected in the optical transmission line using a circulator.  Using this procedure, $M$ delayed replicae of the signal at each WDM channel are produced at each respective output.}}

\section{Results and discussion}

The systems depicted in Figures \ref{fig::dossalidas}, \ref{fig::tdm} and \ref{fig::mz},
for $N=3$ ring CROWs have been analyzed using the FDTD method \cite{meep}.  In each case, independent
simulations have been performed for the subsystems bounded by the dashed-dotted lines in the figures, and the computed transfer functions have been used to obtain the total response.  

The computations have been performed on an equivalent 2D reduced model obtained using the effective index method. 
All the waveguides in our devices are $350$ \si{\nano\metre} wide and an effective index of 
$n_{\text{eff}}=2.4$ has been assumed.  The bent waveguide sections have an inner radious of 
$r=2$ \si{\micro\metre} in all cases. The evanescent couplers have been implemented with a waveguide separation of
$200$ \si{\nano\metre}.  The  
geometry used for the $3$ dB Y-branch couplers and splitters is depicted in Fig. \ref{fig::acoplador}. 
The coupling and splitting responses obtained are very close to ideal 3 dB in all the 
cases.  

Figure \ref{fig::espectros} (a) shows the amplitude responses  and Fig. \ref{fig::espectros} (b) 
the group delays for the $A$ (solid line) and $B$ (dashed line) input/output transfer functions of 
a $N=3$ ring system with $L_c=45$ \si{\micro\metre} 
and constant $L_{\text{off}}=33$ \si{\micro\metre}.  As predicted by the expressions \eqref{eq::FT}, the amplitude 
responses are nearly identical, and the group delay difference between the two outputs is very close to 
a constant value of approximately $2$ \si{\pico\second}.  The variation of the coupling coefficient with the wavelength 
is apparent in the results.

Figure \ref{fig::espectromz} shows the transmission amplitude spectrum for a single input/output 
(dashed line) and the response from the MZ configuration in solid line, showing the FSR doubling effect.  
In this case, $L_c=32.833$ \si{\micro\metre} and $L_{\text{off}}=5.948$ \si{\micro\metre}. 

\section{Conclusion}

In this paper we wave analyzed a CROW chain realized by means of the longitudinal offset technique. The inherent accumulated phase imbalance has been used to propose several architectures for photonic processing based on an alternating chain. We have described the generation of two identically filtered, truly delayed opticals signals, as well as their multiplexing with a controlled time-delay. A Mach-Zender structure which permits to double the FSR has also been presented. All the proposed devices have been proved numerically through 2D FDTD simulation, which accounts for non-ideal effects such as coupler wavelength-dependence and radiation losses, the agreement with the expected results being excellent.  {{Potential applications in optical beamforming antennas and quantum information systems have been discussed.}}

\end{document}